\documentclass[conference, letterpaper, 10 pt]{IEEEtran}

\IEEEoverridecommandlockouts                              

\usepackage{cite}
\usepackage{amsmath,amssymb,amsfonts}
\usepackage{algorithmic}
\usepackage{graphicx}
\usepackage{textcomp}
\usepackage{hyperref}
\usepackage{xcolor}
\usepackage{multicol}
\usepackage{multirow}

\usepackage[caption=false]{subfig}

\usepackage{lipsum} 
\usepackage{booktabs}

\title{\LARGE\bf
Addressing the Intra-class Mode Collapse Problem using Adaptive Input Image Normalization in GAN-based X-ray Images}

\author{Muhammad Muneeb Saad$^{1}$, Mubashir Husain Rehmani$^{2}$ and Ruairi O'Reilly$^{3}$
\thanks{*This research was supported by Risam scholarship award offered by Munster Technological University Cork, Ireland.}
\thanks{$^{1,2,3}$MM. Saad, MH. Rehmani and R. O'Reilly are with Munster Technological University Cork, Ireland. \newline (*Corresponding author to email {\tt\small mmuneeb.saad@gmail.com})
}}

\begin{document}

\maketitle
\thispagestyle{empty}
\pagestyle{empty}

\begin{abstract}

Biomedical image datasets can be imbalanced due to the rarity of targeted diseases. Generative Adversarial Networks play a key role in addressing this imbalance by enabling the generation of synthetic images to augment datasets. It is important to generate synthetic images that incorporate a diverse range of features to accurately represent the distribution of features present in the training imagery. Furthermore, the absence of diverse features in synthetic images can degrade the performance of machine learning classifiers. The mode collapse problem impacts Generative Adversarial Networks' capacity to generate diversified images. Mode collapse comes in two varieties: intra-class and inter-class. In this paper, the intra-class mode collapse problem is investigated, and its subsequent impact on the diversity of synthetic X-ray images is evaluated. This work contributes an empirical demonstration of the benefits of integrating the adaptive input-image normalization for the Deep Convolutional GAN to alleviate the intra-class mode collapse problem. Results demonstrate that the DCGAN with adaptive input-image normalization outperforms DCGAN with un-normalized X-ray images as evident by the superior diversity scores. 
\end{abstract}

\section{INTRODUCTION}

Publicly available biomedical image datasets often contain an insufficient number of images to train a deep learning model \cite{mostapha2019role}. To augment these datasets, Generative Adversarial Networks (GANs) are used to produce synthetic images \cite{wang2021generative}. A GAN consists of two models; the generator for producing synthetic images, and the discriminator, for distinguishing synthetic images from real images. While training a GAN, the generator takes an input of random noise and learns the distribution of real images from feedback provided by the discriminator. The discriminator classifies the generated images as real or synthetic and back-propagates gradient feedback to the generator. The generator updates its learning of feature distributions from discriminator's feedback and endeavours to generate improved synthetic images. 

In biomedical image analysis, diverse features are important for training a classifier to learn the region of interest for better prediction results. Diverse features in biomedical images are more significant than in natural images as biomedical images contain vital information about the disease being classified \cite{lundervold2019overview}. Thus, the GAN must generate diverse images representative of the real biomedical images.

A significant barrier to the generation of diverse images is the mode collapse problem. The mode collapse can occur in one of two forms: the intra-class mode collapse, whereby a GAN generates identical synthetic images from distinct input images for a single class and the inter-class mode collapse, whereby a GAN generates identical synthetic images from distinct input images for all classes. The generator in a GAN can find it difficult to capture every feature from diverse input images to generate synthetic ones. Both variants of mode collapse degrade the performance of GANs for generating diversified images. Subsequently, the performance of machine learning models is degraded when trained on less diversified images \cite{torfi2021evaluation}.

This work proposes the adaptive input-image normalization (AIIN) technique to enable the DCGAN to generate improved diversified synthetic X-ray images. The AIIN is a preprocessing technique for input images that enhances the prominence of desirable input features using a contrast-based computer-vision technique. These features include the shape and texture of body parts in a biomedical image. In the context of X-ray images, these features include the spine, heart, and lungs with their visual signatures like ribs, aortic arch, and distinct curvature of lower lungs. The discriminator learns the input image features more accurately and provides constructive gradient feedback to the generator using normalized X-ray images. Consequently, the generator is forced to produce more diversified images. 

The contribution of this work is to empirically evaluate the efficacy of using AIIN for the DCGAN architecture as means of generating more diversified X-ray images. Key parameters are considered: window size, contrast threshold, and batch size. The AIIN is also compared with other image preprocessing techniques such as median and Gaussian filtering. The occurrence of the intra-class mode collapse problem is evaluated by Multi-scale Structural Similarity Index Measure (MS-SSIM) score. The intra-class diversity is evaluated using Fr\'echet Inception Distance (FID) score.

\section{METHODOLOGY}


In this work, the publicly available dataset published by Kermany et al. \cite{kermany2018} is utilized. The dataset contains 1340 healthy and 3875 Pneumonia chest X-ray images for training purposes. 624 X-ray (234 healthy and 390 Pneumonia) images are available for testing purposes. The dataset is imbalanced, with the healthy chest X-ray images being the minority class and the Pneumonia chest X-ray images being the majority class. The augmentation of healthy chest X-ray images is required to balance the dataset. Images were resized to 128x128 as detailed in \cite{kora2021evaluation}.

\subsection{Data Preprocessing}

The images are preprocessed using AIIN for the DCGAN. As part of the process, contrast-based histogram equalization is used to normalize the images \cite{clahe1994contrast}. Contrast, one of the morphological features, is normalized to highlight the diverse features of the chest X-ray images. Normalized images are visually inspected to find a suitable window size and contrast threshold combination. This work adopts window sizes 4x4, 8x8, and 16x16 based on visual inspection of image features. A window size 32x32 was also considered, but degradation in image quality and feature loss warranted its exclusion. A series of contrast threshold values (0, 5, 10, 20, 50) were selected for normalizing the X-ray images.

The median and Gaussian filtering methods are used to compare the performance of the AIIN technique with alternate image preprocessing approaches. In Gaussian filtering, central pixel value in a window is replaced by weighted average of neighbouring pixels to remove noise in an image. In median filtering, central pixel value of a window is replaced by median of that window. In this work, window size 3x3 and 9x9 were used to normalize the X-ray images as detailed in \cite{shah2020comparative}. A window size 15x15 was also considered but excluded due to the loss of visual information in image.
\begin{figure}[]
    \centering
    \includegraphics[width=0.45\textwidth,height=0.8\textheight,keepaspectratio]{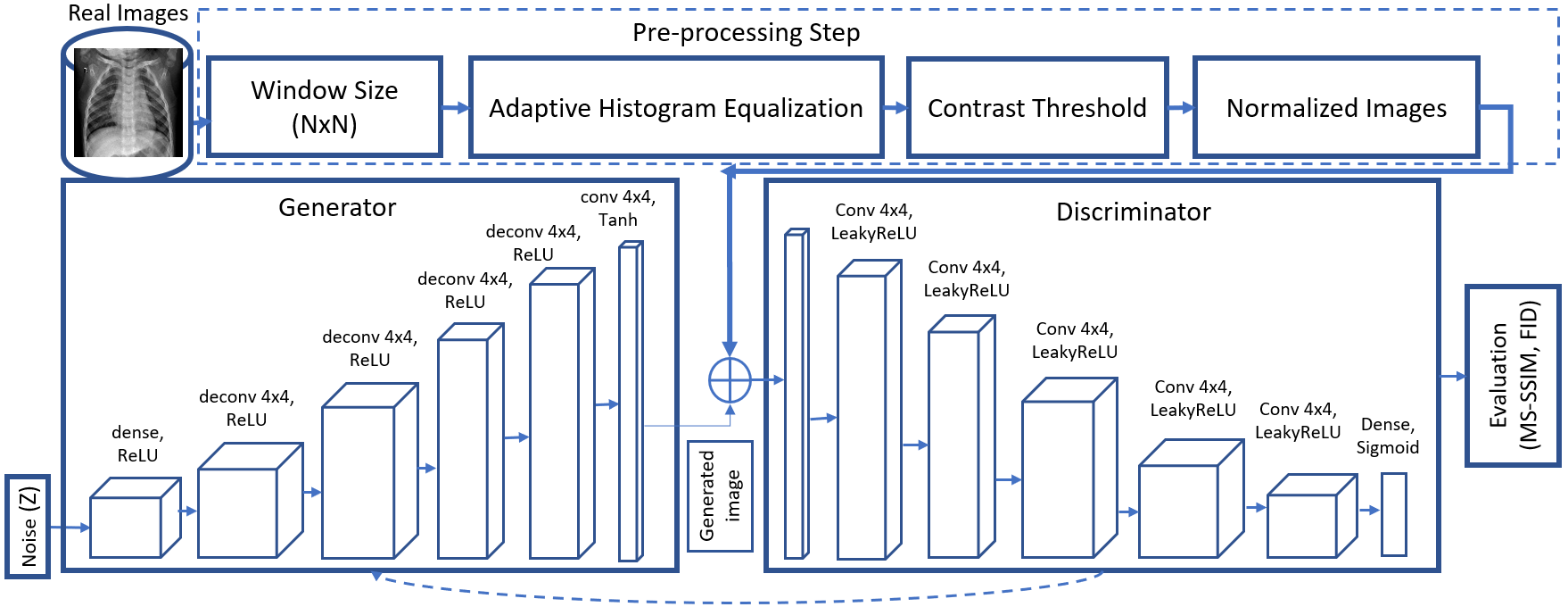}
    \caption{DCGAN Architecture: AIIN is used as a pre-processing step for the DCGAN. MS-SSIM and FID metrics assess the intra-class mode collapse and intra-class diversity of generated X-ray images.}
    \label{Block Diagram}
\end{figure}

\begin{figure*}[]
    \centering
\subfloat[MS-SSIM scores enabling an assessment of the occurrence of intra-class mode collapse.]{%
  \includegraphics[clip,width=0.75\textwidth,height=4.5cm]{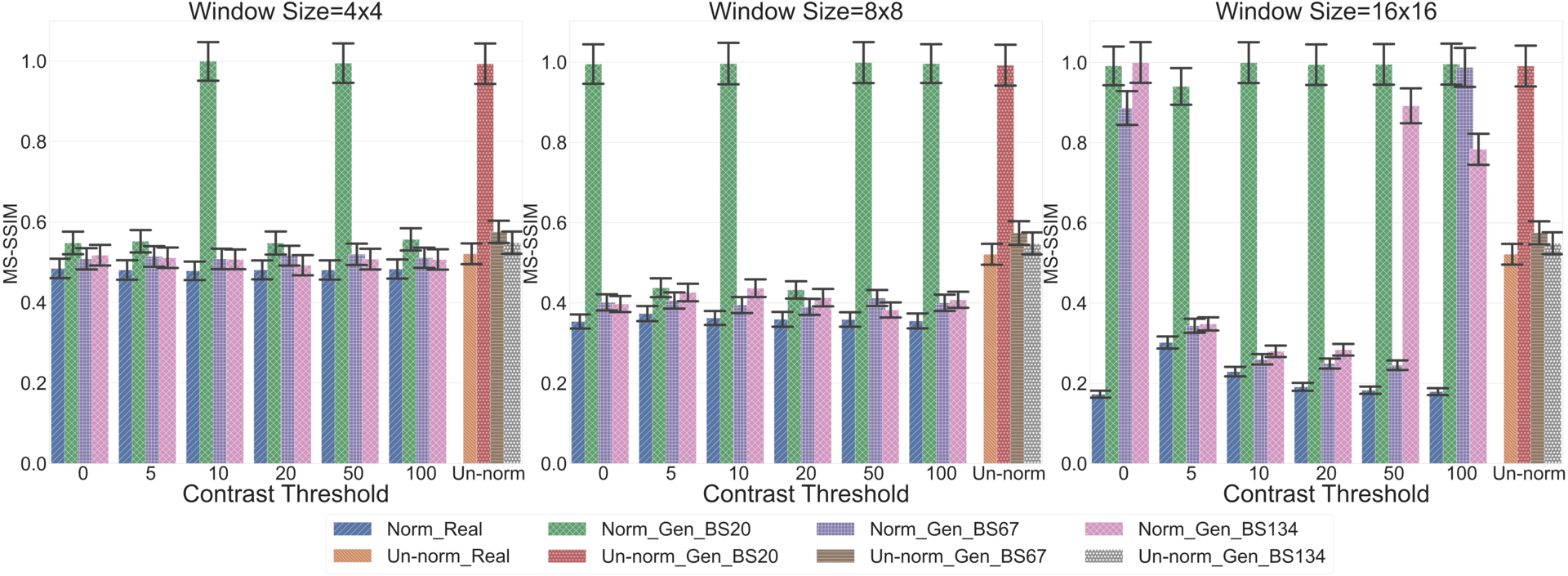}%
}

\subfloat[FID scores enabling an assessment of the level of intra-class diversity.]{%
  \includegraphics[clip,width=0.75\textwidth,height=4.5cm]{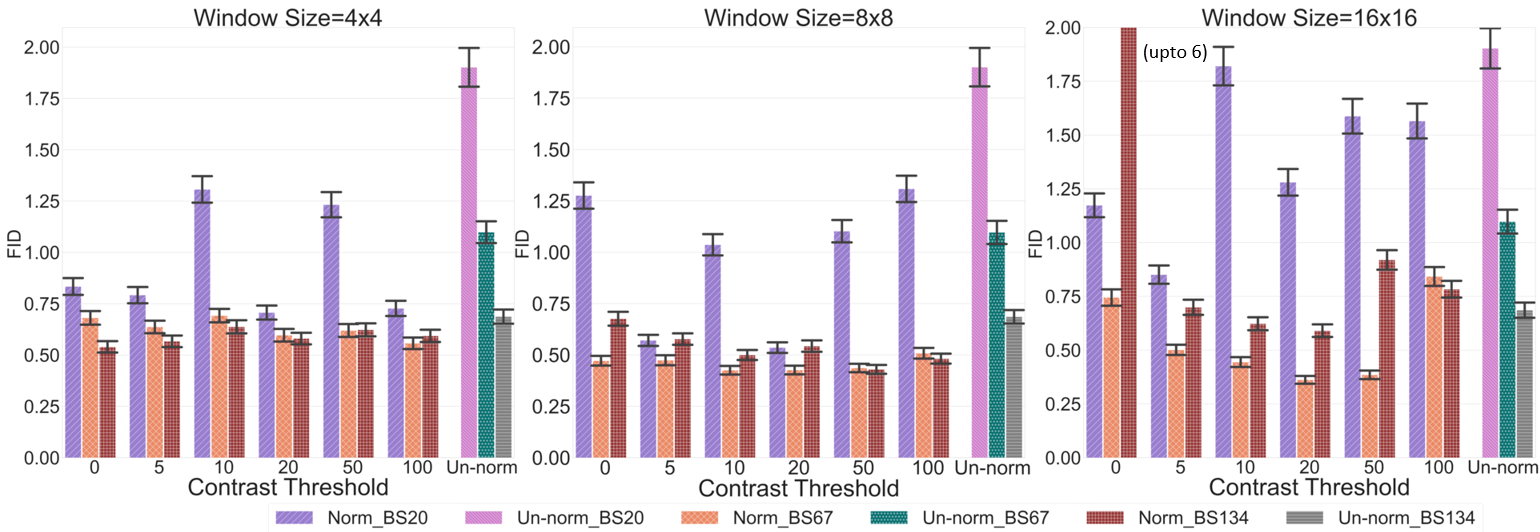}%
}
  \caption{The MS-SSIM and FID scores for the un-normalized and AIIN normalized X-ray images.}
  \label{MS-SSIM_FID_bar}
\end{figure*}

\begin{figure}[]
    \centering
\subfloat[MS-SSIM score and assessment of mode collapse.]{%
  \includegraphics[clip,width=0.4\textwidth,height=1\textheight,keepaspectratio]{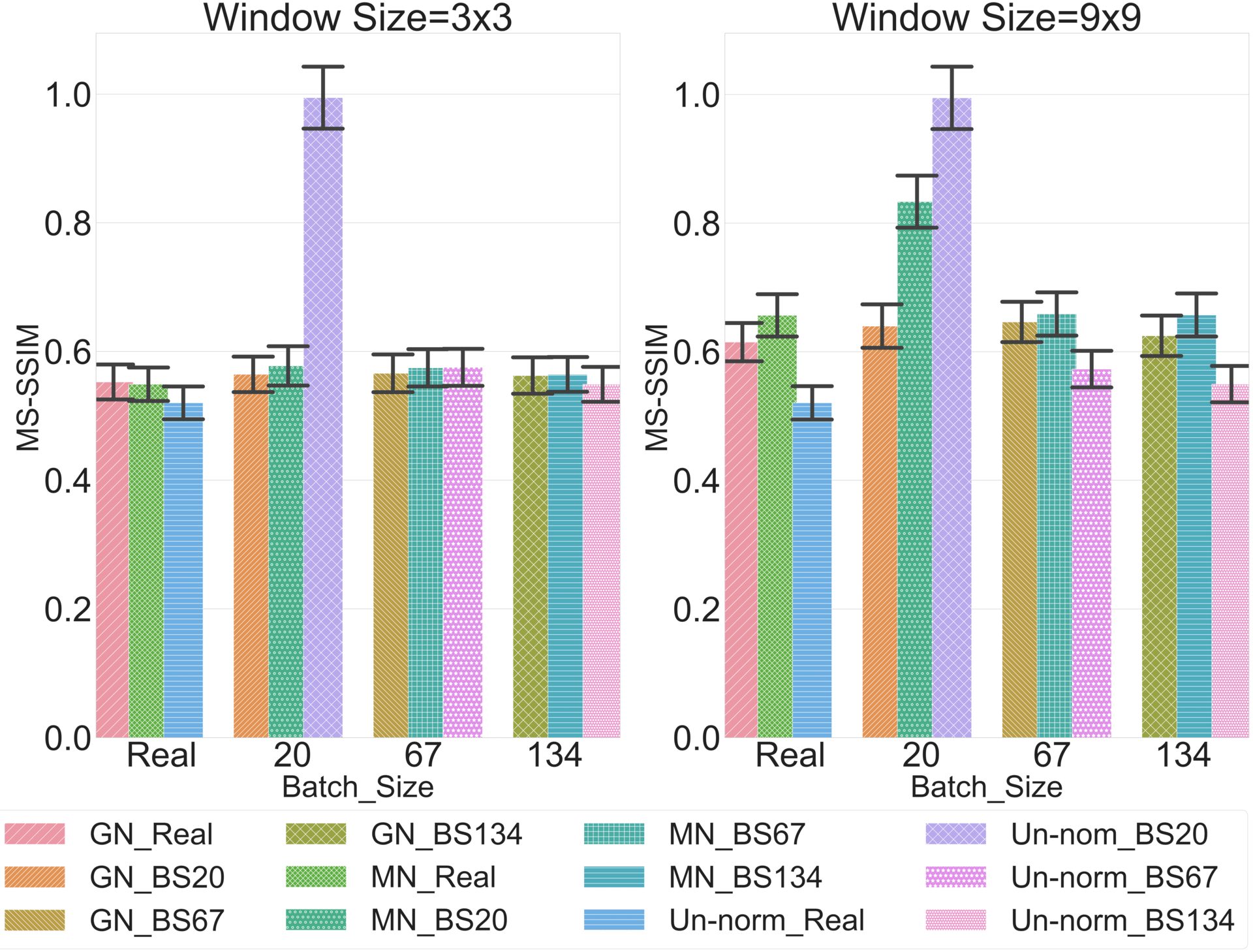}%
}

\subfloat[FID score and the assessment of diversity.]{%
  \includegraphics[clip,width=0.4\textwidth,height=1\textheight,keepaspectratio]{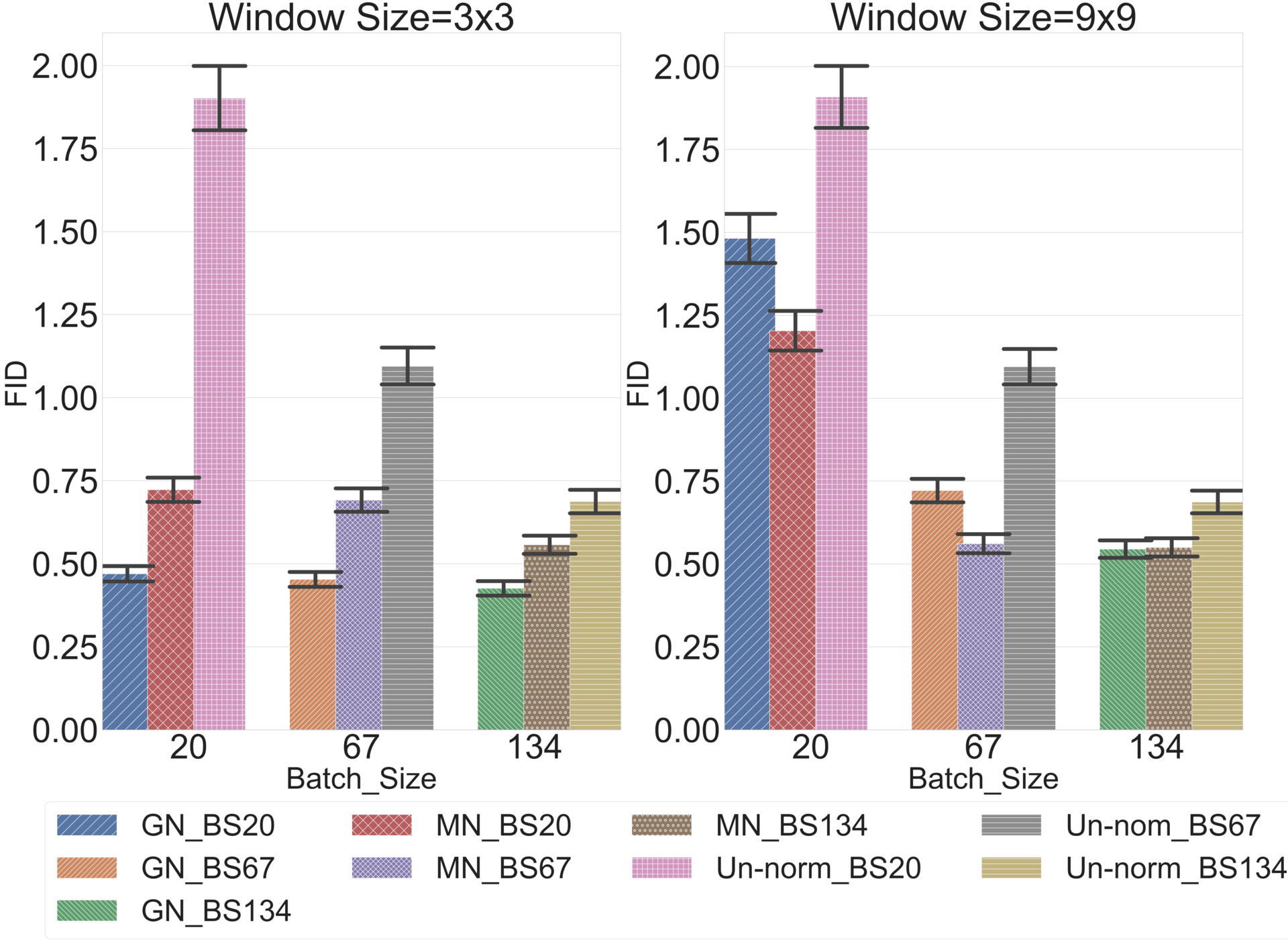}%
}
  \caption{The MS-SSIM and FID scores for the Un-normalized, Gaussian and Mean normalized X-ray images.}
  \label{MS-SSIM_FID_GM}
\end{figure}
\subsection{DCGAN Architecture}

The architecture of the DCGAN is depicted in Fig. \ref{Block Diagram}. The DCGAN \cite{radford2015unsupervised} has been reimplemented and further fine-tuned as detailed in \cite{GANhacks}. The DCGAN uses a Gaussian-latent random input $z$ of 100, ADAM optimizer, and separate real and fake batches for training. A binary cross-entropy loss function is used. The DCGAN is trained for 500 epochs to enable convergence of both models in the GAN \cite{kora2021evaluation}. The DCGAN is trained using the images for each permutation of window size, contrast threshold values, and training batch size. Suitable batch size is considered one that can utilize all of the images available in the minority class. Three different batch sizes that are factors of the available training data for healthy chest X-rays images (20, 67, and 134) were selected to evaluate the DCGAN.
\subsection{Mode Collapse and Diversity of Synthetic Images}

The MS-SSIM score of real and generated images is analyzed to identify the occurrence of mode collapse. FID score is used to identify the level of diversity of synthetic images generated by the DCGAN. These are combined to enable the evaluation of DCGAN's capacity to generate images with a diverse set of features.
\subsubsection{Intra-class Mode Collapse Problem}

Odena et al. \cite{odena2017conditional} first investigated the use of MS-SSIM to measure the intra-class diversity of generated imagery and assess the occurrence of intra-class mode collapse in GANs. The similarity between two images is computed based on image pixels and structures. MS-SSIM scores are measured between randomly selected pairs of real-to-real images and pairs of synthetic-to-synthetic images separately, with the cumulative mean score being reported. The range of the MS-SSIM score lies between 0 and 1. In this work, 670 image pairs are used randomly to measure the MS-SSIM score. A higher score for synthetic images as compared to real images is indicative of mode collapse occurring. Synthetic images should possess a similar or lower MS-SSIM score compared to real images.

MS-SSIM is computed between two image samples, x and y as defined in Eq. \eqref{Eq. ms-ssim} \cite{borji2019pros}.
\begin{equation}
\resizebox{0.9\hsize}{!}{$%
\operatorname{MS}-\operatorname{SSIM}(x, y)=I_{M}(x, y)^{\alpha_{M}} \prod_{j=1}^{M} C_{j}(x, y)^{\beta_{j}} S_{j}(x, y)^{\gamma_{j}}\label{Eq. ms-ssim}
$}%
\end{equation}
In Eq. \eqref{Eq. ms-ssim}, contrast (C) and structure (S) features are computed with scale $j$. Luminance (I) is computed at the coarsest scale denoted by M. The $\alpha$, $\beta$, and $\gamma$ are the weight parameters as reported in \cite{wang2003multiscale}.
\subsubsection{Intra-class Diversity}

The intra-class diversity of generated images is assessed using FID. FID evaluates the distance between synthetic images and real images using feature activations \cite{miyato2018cgans}. In this work, FID is measured using a sample size of 1340 images separately selected from real and synthetic images with a score ranging from 0.0 to $+\infty$. A lower FID score indicates a higher degree of diversity of synthetic images related to real images \cite{borji2019pros}.
\begin{equation}
\resizebox{0.9\hsize}{!}{$%
F I D(r, s)=\left\|\mu_{r}-\mu_{s}\right\|_{2}^{2}+\operatorname{Tr}\left(\Sigma_{r}+\Sigma_{s}-2\left(\Sigma_{r} \Sigma_{s}\right)^{\frac{1}{2}}\right)\label{FID}
$}%
\end{equation}
In Eq.\eqref{FID}, $r$ and $s$ denote real and synthetic images while $\left(\mu_{r}, \Sigma_{r}\right)$ and $\left(\mu_{s}, \Sigma_{s}\right)$ denote their mean and covariances. Whereas Tr denotes the trace of matrices. FID uses the last pooling layer of the Inception V3 model, which contains a 2048 dimensional feature; it requires 2048 or more training image samples as input. As there are only 1340 healthy chest X-ray images available, a Pre-aux Classifier layer containing 768-dimensional features is used instead of the 2048 dimensional feature.
\subsection{Assessing the Utility of Synthetic Chest X-ray Images}

To assess the utility of synthetic X-ray images, a sequential CNN \cite{siddiqi2019automated} is implemented for the classification of healthy X-ray images. The intent is to augment the minority class with synthetic X-ray images with varying degrees of similarity and diversity. This will enable an evaluation of the efficacy of AIIN for the DCGAN in augmenting chest X-ray images. Whereas the CNN was used to classify Pneumonia X-ray images in \cite{siddiqi2019automated} as the results were assessed for Pneumonia representing positive labels.

Synthetic images were rescaled to 150x150 with the Open-CV library, and the CNN model reimplemented as detailed in \cite{siddiqi2019automated}. The CNN model is trained on the dataset with 13 GAN-based augmentation variants. 1340 synthetic images are generated for each variant and used to address the data imbalance problem. Selection of variants was based on those with the most promising scores for MS-SSIM and FID. Geographical transformation such as rotation 15\textdegree, shear, and zoom range of 0.2 was also used. Classification scores are compared with the un-normalized generated healthy chest X-ray images for all instances as detailed in Table \ref{tabclasscompare}.
\section{Results and Discussion}

The MS-SSIM and FID scores of AIIN, Gaussian, and Mean normalized with un-normalized generated X-ray images are depicted in Fig. \ref{MS-SSIM_FID_bar} and Fig. \ref{MS-SSIM_FID_GM} respectively.

The intra-class mode collapse is identified by a higher MS-SSIM score of un-normalized synthetic X-ray images than real images. The AIIN has alleviated the mode collapse by improving the capacity of DCGAN to generate diversified normalized X-ray images, as indicated by the improved MS-SSIM scores. Parameters like window size, contrast threshold, and batch size significantly impact the generation of diversified synthetic images, as indicated by the varying MS-SSIM scores. Gaussian and median filtering approaches achieve higher MS-SSIM scores of normalized synthetic images than AIIN. These filtering approaches suppress noise in the image but blur the edges of an image, reducing features' structural information yet improving the MS-SSIM score. Therefore, these filtering approaches have no advantage to DCGAN for alleviating the mode collapse problem.

FID analysis indicates the efficacy of AIIN in improving the intra-class diversity of synthetic normalized X-ray images as depicted in Fig. \ref{MS-SSIM_FID_bar}(b). Results show that the parameters: window size, contrast threshold, and batch size have a considerable impact on the diversity of synthetic X-ray images, as indicated by the varying FID scores. The AIIN achieves relatively better FID scores than Gaussian and median filtering techniques, as depicted in Fig. \ref{MS-SSIM_FID_GM}(b). 

Several GAN-based variants have been implemented to augment healthy X-ray images, as denoted in Table \ref{tabclasscompare}. First row in Table \ref{tabclasscompare} details the CNN results reported in \cite{siddiqi2019automated} while the second row shows the reimplemntation of results. The predictions for minority class containing healthy X-ray images represent negative labels as the CNN is a classifier of Pneumonia X-ray images. Therefore, the results are evaluated by the specificity score. The best specificity score is achieved at batch size 134 using window sizes 4x4 and 8x8 with a contrast threshold of 0 and 50. This improvement demonstrates the normalized synthetic images' capacity to augment the healthy X-ray images outperforming the un-normalized approaches. The CNN has issues like randomization of features and overfitting. In this case, the CNN focuses on classifying Pneumonia and learning the features of lung segments to differentiate healthy images from Pneumonia images in the chest X-ray images. The lung segment is highlighted more using 4x4 and 8x8 window sizes as compared to the alternate permutations of the experiment yet achieves good classification measures. Compared with Gaussian and median filtering approaches, the AIIN demonstrates advantages in that it does not degrade the structural information of features in images yet achieves better classification scores, as presented in Table. \ref{tabclasscompare}.
\section{CONCLUSIONS}

This work proposes an AIIN technique for the DCGAN to improve the diversity of generated chest X-ray images. Results show that the DCGAN with AIIN can generate more diversified X-ray images than DCGAN without AIIN (indicated by better MS-SSIM and FID scores) while alleviating the mode collapse problem. The AIIN also performed better than the Gaussian and median filtering preprocessing techniques via diverse image features. Furthermore, the efficacy of the proposed approach is verified by using the augmented (generated X-ray images combined with the real images) images to train machine learning classifiers.

\begin{table}[hbt!]
\centering
\caption{CNN's classification performance of Pneumonia vs Healthy X-ray images under different GAN-based training scenarios.}
      \setlength{\tabcolsep}{3pt}
      \resizebox{.4\textwidth}{!}{%
      \begin{tabular}{llllllllll}
       \toprule
       \textbf{Aug} & \textbf{BS} & \textbf{WS} & \textbf{CT} & \textbf{MS-S.} & \textbf{FID} & \textbf{Acc.} & \textbf{Prec.} & \textbf{Rec.} & \textbf{Spec.} \\ 
       \midrule
       N/A & N/A& N/A& N/A& N/A& N/A& 94.39 & 0.92 & 0.99 & 0.86 \\
       N/A & N/A& N/A& N/A& N/A& N/A& 91.20 & 0.88 & 0.99 & 0.78 \\
       Un-N. & 20 & N/A& N/A& +0.473 & 1.903 & 87.5 & 0.84 & 0.98 & 0.69 \\
       Un-N. & 67 & N/A& N/A& +0.054 & 1.096 & 87.20 & 0.84 & 0.98 & 0.69 \\
       Un-N. & 134 & N/A& N/A& +0.029 & 0.687 & 88.94 & 0.86 & 0.98 & 0.74 \\
       AIIN & 134 & 4x4 & 20 & \textbf{+0.013} & 0.580 & 87.6 & 0.85 & 0.98 & 0.71 \\
       AIIN & 134 & 8x8 & 50 & \textbf{+0.025} & 0.430 & \textbf{91.50} & 0.89 & 0.99 & \textbf{0.79} \\
       AIIN & 67 & 16x16 & 10 & \textbf{+0.031} & 0.444 & 87.98 & 0.85 & 0.98 & 0.71 \\
       AIIN & 134 & 4x4 & 0 & +0.036 & \textbf{0.540} & \textbf{91.50} & 0.89 & 0.99 & \textbf{0.79} \\
       AIIN & 67 & 8x8 & 10 & +0.035 & \textbf{0.425} & 90.06 & 0.87 & 0.99 & 0.74 \\
       AIIN & 67 & 16x16 & 20 & +0.058 &\textbf{0.362} & 84.13 & 0.82 & 0.96 & 0.65 \\
       GN & 134 & 3x3 & N/A& +0.01 & 0.426 & 85.57 & 0.82 & 0.99 & 0.63 \\
       GN & 134 & 9x9 & N/A& +0.0008 & 0.547 & 90.38 & 0.88 & 0.99 & 0.76 \\
       MN & 134 & 3x3 & N/A& +0.015 & 0.557 & 88.62 & 0.85 & 0.99 & 0.72 \\
       MN & 134 & 9x9 & N/A& +0.001 & 0.550 & 87.50 & 0.84 & 0.99 & 0.69 \\
       \bottomrule
       \multicolumn{10}{l}{Aug:Augmentation; Un-N:Un-normalized; AIIN:AIIN-Normalized}\\
       \multicolumn{10}{l}{GN:Gaussian-Normalized; MN:Median-Normalized; BS:Batch Size}\\
       \multicolumn{10}{l}{WS:Window Size; CT:Contrast Threshold; MS-S:MS-SSIM}\\
       \multicolumn{10}{l}{Acc:Accuracy; Prec:Precision; Rec:Recall; Spec:Specificity}
       \end{tabular}}
       \label{tabclasscompare}
\end{table}

\bibliographystyle{IEEEtran}



\end{document}